\documentclass[english,keywords,aps,amsmath,amssymb,twocolumn]{revtex4-1}
\usepackage[T1]{fontenc}
\usepackage[latin1]{inputenc}
\usepackage{babel}
\usepackage{amssymb}
\usepackage{graphicx}
\usepackage{color}
\usepackage{bm}
\usepackage{longtable}
\usepackage{amsmath} 
\usepackage{amsfonts}
\usepackage{amssymb}
\begin{document}
\title{Semi-device independent randomness certification using Mermin's proof of Kochen-Specker contextuality}
\author{ A. K. Pan }
\email{akp@nitp.ac.in}
\affiliation{National Institute of Technology Patna, Ashok Rajpath, Patna, Bihar 800005, India}
\begin{abstract}
Randomness is a potential resource for cryptography, simulations and algorithms. Non-local correlations violating Bell's inequality certify the generation of bit strings whose randomness is guaranteed in a device-independent manner. We provide interesting semi-device independent randomness certification protocols by Kochen-Specker (KS) contextuality. For this, we first cast the Mermin's magic-square proof of KS contextuality for two-qubit system as a semi-device independent communication game in prepare-measure scenario. This provides a semi-device independent certification of two-bit of randomness. Further, by using Mermin's magic-star proof of KS theorem involving three-qubit system, we certify three-bit of randomness. We conjecture that our proposals can be extended to certify any arbitrary bit of randomness through a suitable KS proof of contextuality valid for higher dimensional system.
\end{abstract}
\maketitle

\section{Introduction}
Randomness is a powerful resource having a wide field of applicability ranging from scientific research to our daily life. True randomness implies that one cannot deduce a pattern even though every piece of knowledge of the random number generator is accessible along with a significant part of the random sequence.  Classical algorithms, whatever powerful it may be, can only produce a pseudo-random number, whose unpredictability relies on the complexity of the generator \cite{matsumoto}.  On the other hand, the quantum theory provides intrinsic randomness through the unpredictability of the Born rule. For example, an unpredictable sequence of random bits can be generated when a photon passes through a suitable beam-splitter or a spin-polarized neutral particle passes through a Stern-Gerlach setup.  Such unpredictability cannot be given epistemological status but is intrinsic to the quantum theory.

Practically, such randomness relies on one's trust in the functioning of the quantum device. If one is skeptical, as one has the right to be if she buys the device, it is impossible for her to distinguish whether the generated random bit comes from a pre-programmed classical random number generator or not.  How can then the seller convince the buyer that the randomness generated by a device is caused by nature itself and not from our lack of knowledge about the complexity of the device?  Clearly, this way of generating randomness can never be trusted, as one can always design a classical device (so-called local or non-contextual deterministic model) that can mimic the outputs. It is thus desirable if the generated randomness can be certified solely by statistical tests of the inputs and outputs without requiring the characterization of the devices. Bell's theorem \cite{bell, brunnerreview}(all quantum statistics cannot be reproduced by local models) provides such a device-independent certification, in that no characterization of devices are assumed and are just treated as black boxes \cite{acin07, brunnerreview}. The device-independent randomness generation relies on a fundamental relation between the non-locality of quantum theory and its random character. 

Such a strategy of certifying device-independent randomness was first put forwarded by Colbeck \cite{colbeck06} in his PhD thesis.  Adopting the similar strategy in \cite{pironio} the relation between randomness and violation of Bell's inequality is established through the non-local guessing games. Alice and Bob hold input $x\in \{1,2,...n\}$ and $y\in \{1,2,...m\}$ respectively  and each of the inputs produces dichtomic outputs $a\in \{1,-1\}$ and $b=\in \{1,-1\}$.  The joint probability $P(a b|x,y)$ can be obtained when Alice and Bob perform measurements according to the given inputs. In the non-local guessing game \cite{nieto} there is another party, Eve, whose goal is to guess the Alice's outcome with highest possible probability. A strategy of Eve can be that she prepares the quantum state $|\Psi_{ABE}\rangle\in \mathcal{H}_{A}\otimes \mathcal{H}_{B}\otimes \mathcal{H}_{E}$ for Alice and Bob, so that  $\rho_{AB}$ can be obtained by tracing out her system. Given inputs $x$ and $y$, Alice and Bob measures a set of positive operator valued measures (POVMs) $\{A_{a|x}\}$ and $\{B_{b|y}\}$ respectively. Thus, $P(a b|x,y)=Tr[(A_{a|x}\otimes B_{b|y}\otimes \mathbb{I})\rho_{ABE}]$. Given an input $x^{\ast}$, the local guessing probability can be written as $	P(a|x^{\ast})=max_{a}Tr[(A_{a|x^{\ast}}\otimes \mathbb{I}\otimes F_{a})\rho_{ABE}]$ where $\{F_a\}$ is the Eve's set of measurement operators. It is established in \cite{acin07} that $P(a|x^{\ast})=f(\mathcal{\beta})$ where $\beta$ is a Bell expression. The min-entropy can be used as a measure of randomness, so that, $H_{\infty}(a|x)=-log_{2} 	P(a|x^{\ast})$ which provides the device-independent certified randomness in bits.

Since then a flurry of works have been reported \cite{colbeck, Gallego,acin12, curchod,Wei,acin16,and18} and verified experimentally \cite{Yang,Peter}. However, the device-independent randomness certification faces practical challenges arises due to the requirement of loophole-free violation of Bell inequality by lowering the bit rate.  In recent times, loophole-free Bell tests have been realized \cite{loopholefree} which in turn enables experimental demonstrations of device-independent random number certification \cite{Yang, Peter}.  However, due to the lack of finite-data efficiency, random number generation protocol based on the Bell test requires a very large number of trials with loophole-free Bell tests. Further, the device-independent certification of random number generator in a prepare-measure scenario is proposed in \cite{lungi}. Semi-device-independent randomness certification protocols \cite{li11,li12,Wen,pas,van,brask,ioa} in the prepare-measure scenario have also been proposed where the dimension of the quantum system is known. 

In a bipartite Bell scenario, by considering the arbitrary number of sequential measurements at one of the two local systems an unbounded amount of certified random bits was generated \cite{curchod}. Let a measurement Pauli observable $\sigma_{x}$ is performed on a qubit state $|0\rangle$ which produces random outputs. If a non-commuting observable $\sigma_{z}$ is measured in sequence, the output will be more random. One can then sequentially perform the measurements of these two observables for arbitrary number of times and such a chain of measurements can then generate an unbounded amount of randomness. However, such randomness for the qubit system is not certified unless the system is entangled with another system and provides Bell's inequality violation. This is exactly done in \cite{curchod} to certify the unbounded amount of randomness. Note here that, in \cite{curchod}, in order to get the violation of Bell's inequality for an arbitrary number of sequential Bobs, a residual entanglement has to be present after each measurement of Bob. This means that each sequential Bob has to perform unsharp measurement so that the entangled state is minimally disturbed but enough to violate Bell's inequality. 

In contrast to the aforementioned scenario, we demonstrate here that if one considers a single system having $dim(H)\geq 3$ and performs suitable sequential projective measurements of commuting observables, the generated randomness can be semi-device independently certified by Kochen-Specker (KS) quantum contexuality \cite{kochen}.  The original KS theorem \cite{kochen} was demonstrated using 117 projectors for a single qutrit system. Later, simpler versions of it using a lower number of projectors have been provided \cite{ker}. Instead of projectors, using dichotomic observables, interesting versions of KS theorem were proposed by Peres \cite {peres} which was further extended to the state-independent form by  Mermin\cite {mermin}.  These proofs are converted into testable inequalities, valid for any KS noncontextual model, but violated by quantum mechanics \cite{cabello08,pan}. The notion of contextuality has also been extensively studied both theoretically \cite{theor,kly} and experimentally \cite{hasegawa,nature,liu,ams,radek}. 
 
In this work, we propose semi-device independent certification of two-bit of  randomness through Mermin's \cite{mermin} magic-square proof of KS contextuality which requires at least two-qubit system. This is  done by succinctly casting Marmin's \cite{mermin} magic-square proof of KS theorem as new communication game in prepare-measure scenario. Note that the semi-device independent  randomness certification in prepare-measure scenario has been studied in many earlier works \cite{li11,li12,Wen,pas,van,brask,ioa}. We use two general assumptions; the bounded dimension and no information leakage from the devices which are uncharacterized. Similar to the many other prepare-measure games in the literature, our scheme also does not require entanglement and certification of our protocol is guaranteed by KS contextuality. Further, we demonstrate that three-bit of randomness can be certified by using Mermin's magic-star proof that requires at least  three-qubit system. Note that, two or more bit of randomness for the two-qubit system was generated in \cite {acin16,and18,curchod} (certified by quantum non-locality) uses non-projective measurements. Here we certify two-bit (three-bit) of randomness for a single two-qubit (three-qubit) system using sequential projective measurements of three (four) commuting observables, suitable for exhibiting quantum contextuality.  It is also discussed that magic-square proof of KS contextuality can be seen as a proof of preparation contextuality to explain that a maximally mixed state $\mathbb{I}/d$ prepared by distinct procedures in quantum theory cannot be represented equivalently in an ontological model.  We conjecture that for suitable Mermin-type proof of KS contextuality for a higher-dimensional system, an arbitrary bit of randomness can be certified.

%This paper is organized as follows. In Sec. II, we briefly recapitulate the essence of Mermin's magic star proof from the perspective of Spekkens \cite{spekk05} generalized formulation of non-contextual ontological model. We then provided a communication game in Sec. III, whose success probability is solely dependent on the non-contextual inequality originated from magic-square proof. We further show how two bit of randomness can be generated from Mermin's magis-square proof of KS contextuality. We extended the formulation to certify three bit of randomness by using Mermin's magic star proof in Sec. IV. A brief summary and discussion is placed in Sec. V.   

\section{Mermins' magic-square proof of KS theorem revisited}
We invoke an elegant framework of an ontological model \cite{hari,spekk05} of quantum mechanics to introduce the KS contextuality from modern perspective.   In quantum mechanics, a preparation procedure $(P)$ produces a density matrix $\rho$ and measurement procedure $(M)$ (in general described by a suitable POVM $(E_k$) provides the probability of a particular outcome $ k $ is given by $p(k|P, M)=Tr[\rho E_{k}]$, which is the Born rule. In an ontological model of quantum theory, it is assumed that whenever $\rho$ is prepared by $P$, a probability distribution $\mu_{P}(\lambda|\rho)$ in the ontic space is prepared, satisfying $\int _\Lambda \mu_{P}(\lambda|\rho)d\lambda=1$ where $\lambda \in \Lambda$ and $\Lambda$ is the ontic state space. The probability of obtaining an outcome $k$ is given by a response function $\xi_{M}(k|\lambda, E_{k}) $ satisfying $\sum_{k}\xi_{M}(k|\lambda, E_{k})=1$ where a measurement operator $E_{k}$ is realized through $M$. A viable ontological model should reproduce the Born rule, i.e., $\forall \rho $, $\forall E_{k}$ and $\forall k$, $\int _\Lambda \mu_{P}(\lambda|\rho) \xi_{M}(k|\lambda, E_{k}) d\lambda =Tr[\rho E_{k}]$.

An ontological model of quantum theory can be assumed to be non-contextual in the following way \cite{spekk05};  if two experimental procedures are operationally equivalent then they have equivalent representations in the ontological model.   An ontological model is assumed to be measurement non-contextual if $\forall P \ \forall k:  p(k|P, M)=p(k|P, M^{\prime})	\Rightarrow \xi_{M}(k|\lambda, E_{k})=\xi_{M^{\prime}}(k|\lambda, E_{k})$  is satisfied, where the POVM $E_k$  is realized by two distinct measurement procedures $ M $ and $ M' $. Along the same line, an ontological model is assumed to be preparation non-contextual if $\forall M  \ \forall k:  p(k|P, M)=p(k|P^{\prime},M)	\Rightarrow \mu_{P}(\lambda|\rho)=\mu_{P^{\prime}}(\lambda|\rho)$ where the $\rho$ is prepared by two distinct preparation procedures $ P $ and $ P^{\prime}$ \cite{spekk05, pan19}.

The KS non-contextuality assumes the aforementioned measurement non-contextuality along with the outcome determinism for the sharp measurement, i.e., $\xi(k| P_{k}, \lambda)\in \{0,1\}$ where $P_{k}$ is a projector. A dichotomic observable $M$ can always be written as $M=2P_{k}-\mathbb{I}$ leading $\xi(k| M, \lambda)\in \{-1,1\}$. In Mermin's proof \cite{mermin} of KS contextuality,  the product rule is needed to be invoked which is a direct consequence of measurement non-contextuality assumption. For the joint measurement of two mutually commuting observables (say, $\widehat{M}_{1}$ and $\widehat {M}_2$), the notion of measurement non-contextuality of an ontological model is characterized by the following feature (the product rule), i.e.,  $\xi(m_1,m_2|M_{1}, M_{2},\lambda)=\xi(m_1|M_1,\lambda)\xi(m_2|M_2,\lambda)$. 

Mermin's magic-square proof \cite{mermin,cabello08,pan} of KS contextuality requires nine two-qubit observables are the following.
\begin{equation}
\label{pmarray}
\begin{array}{clrr}     
C_1=\sigma_{y}\otimes \sigma_{z} & \hskip 0.2cm B_1=\sigma_{z}\otimes \sigma_{y} &\hskip 0.3cm   A_1={\sigma_{x}\otimes\sigma_{x}} \\
\\
A_2=\sigma_{z} \otimes \sigma_{x}  &\hskip 0.2cm  C_2=  \sigma_{x}\otimes \sigma_{z}  &  B_2=\sigma_{y}\otimes\sigma_{y}\\
\\
B_3=\sigma_{x}\otimes\sigma_{y} &\hskip 0.2cm  A_3=\sigma_{y}\otimes\sigma_{x}& C_3=\sigma_{z}\otimes\sigma_{z}
\end{array}
\end{equation}
Note that, the observables in each row and each column  of the array given by Eq.(2) are mutually commuting and \emph{any} quantum state $\rho\in \mathbb{C}^2\otimes\mathbb{C}^2$, is the eigenstate of the product of the three observables in each row ($R_{i=1,2,3}$) and in each column($L_{i=1,2,3}$), satisfies the relation $\langle \rho R_1 \rangle=\langle \rho R_2 \rangle=\langle \rho R_3\rangle=-\langle \rho L_1\rangle=-\langle \rho L_2\rangle=-\langle \rho L_3\rangle\equiv\langle \rho (\mathbb{I}_{2}\otimes \mathbb{I}_{2})\rangle=1$. 

By using the product rule the response functions for $R_{i=1,2,3}$ and ($L_{i=1,2,3}$) can be written as 
\begin{eqnarray}
\nonumber
&&\xi(R_1)= \xi_{R_1}(C_1) \  \xi_{R_1}(B_1) \ \xi_{R_1}(A_1) =1\\
\nonumber
&&\xi(R_2)= \xi_{R_2}(A_2) \  \xi_{R_2}(C_2) \ \xi_{R_2}(B_2) =1\\
\nonumber
&&\xi(R_3)= \xi_{R_3}(B_3) \  \xi_{R_3}(A_3) \ \xi_{R_3}(C_3) =1\\
\label{prime1}
&&\xi(L_1)= \xi_{L_1}(C_1) \  \xi_{L_1}(A_2) \ \xi_{L_1}(B_3) =-1\\
\nonumber
&&\xi(L_2)= \xi_{L_2}(B_1) \  \xi_{L_2}(C_2) \ \xi_{L_2}(A_3) =-1\\
\nonumber
&&\xi(L_3)= \xi_{L_3}(A_1) \  \xi_{L_3}(B_2) \ \xi_{L_3}(C_3) =-1
\end{eqnarray} 
where we use the shorthand notation $\int\xi(k|R_1,\lambda)\mu(\lambda,\rho)d\lambda\equiv\xi(R_{1})$ and similarly for others. Also, $\xi_{R_1}(C_1)$ and $\xi_{L_1}(C_1)$ denote the response function $\xi(C_1)$ in two contexts $R_1$ and $C_1$ respectively. In a KS non-contextual  model  $\xi_{R_1}(C_1)=\xi_{L_1}(C_1)\equiv \xi(C_1) \in \{-1,1\}$.

Since every response function appears twice in Eq. (\ref{prime1}), using KS non-contextual value assignments,  product of all $\xi(R_{i=1,2,3})$ and $\xi(L_{i=1,2,3})$ yields $+1$. But, quantum mechanics predicts $-1$, thereby contradicting the KS non-contextuality. Note also that, five out of six relations in Eq. (\ref{prime1}) can be consistent with KS value assignments.   

It order to empirically test the above logical incompatibility, non-contextual inequality was proposed in \cite{cabello08,pan} 
\begin{equation}
\label{newmc}
\left\langle\Delta\right\rangle_{ks} = \left\langle R_{1}\right\rangle +\left\langle R_{2}\right\rangle+\left\langle R_{3}\right\rangle -\left\langle L_{1}\right\rangle-\left\langle L_{2}\right\rangle-\left\langle L_{3}\right\rangle\leq 4
\end{equation}
Quantum mechanics violates the inequality in Eq. (\ref{newmc}) by predicting $\left\langle\Delta\right\rangle_{QM}=6$ for any state in two-qubit systems.

\section{A communication game based on Magic-square proof and randomness certification}
\begin{figure}[ht]
\centering
\label{fig}
\includegraphics[width=1\linewidth]{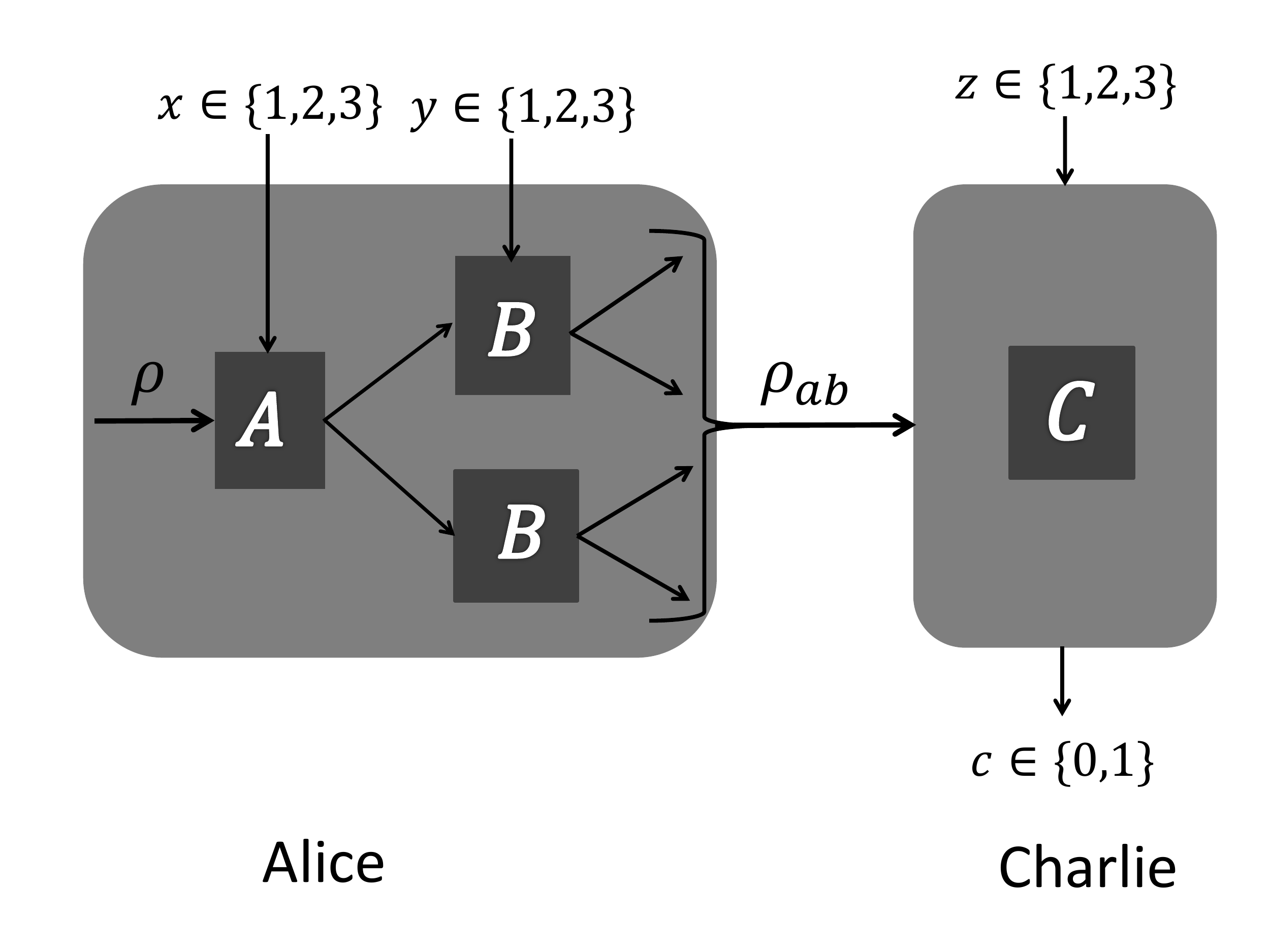}
\caption{The schematic of semi-device independent prepare-measure communication game implementing Mermin's proof of Kochen-Specker contextuality. Alice prepares the particle in quantum state $\rho_{ab}$ by sequential measurements of two commuting observables upon receiving the inputs ($x,y \in \{1,2,3\}$) and sends it to Charlie who performs the measurement according to his inputs $z\in \{1,2,3\}$.}
\end{figure}
   We  propose a new communication game in  so-called prepare-measure scenario \cite{li11,li12,Wen,pas,van,brask,ioa,lungi} where the success probability can be shown to be solely dependent on Mermin's magic-square inequality in Eq. (\ref{newmc}). This is a semi-device independent approach that requires two uncharacterized devices - one for state preparation for Alice and another one for measurement for Bob.  However, only a few general assumptions are made, such as the quantum system is of bounded dimension and no leakage of information from the devices. Such a semi-device independent random generation protocol has already been considered in many earlier works \cite{li11,li12,Wen,pas,van,brask,ioa} without using entanglement. In our semi-device independent prepare-measure scenario, the randomness is certified by KS contextuality. 

In our prepare-measure game, given a quantum state $\rho \in \mathbb{C}^{2}\otimes \mathbb{C}^{2}$, upon receiving inputs $x\in \{1,2,3\}$  followed by $y\in \{1,2,3\}$ (having outputs  $a\in \{0,1\}$ and $b\in \{0,1\}$ respectively), Alice prepares quantum state $\rho_{ab}$ by performing the sequential measurements of two two-qubit commuting observables.   Charlie performs the measurements on the particles received from Alice according to the inputs $z\in \{1,2,3\}$ producing outputs $c\in \{0,1\}$. The scenario is depicted in Fig. 1.  The winning condition of the game is the following;  if $x=y=z$   then $a \oplus_{2} b \oplus_{2} c =0$ and if $y=x\oplus_{3} 1, z=x\oplus_{3} 2$ then $a \oplus_{2} b \oplus_{2} c \neq 0$. Here  $\oplus_{2}$ and $\oplus_{3}$ are the addition modulo $2$ and $3$ respectively. 

The winning probability of the game can then be written as
\begin{eqnarray}
\label{sucprob1}
\mathcal{P}_{w}= &&\frac{1}{6}\sum\limits_{x,y,z=1}^{3}\Big[P(a \oplus_{2} b \oplus_{2} c =0|x=y=z) \\
	\nonumber
	&+&  P(a \oplus_{2} b \oplus_{2} c = 1|x, y=x\oplus_{3}\ 1; z=x\oplus_{3}\ 2) \Big]
	\end{eqnarray}
Given the winning condition, it is straightforward to figure out from Eq. (\ref{prime1})  that at most five out of six conditions can be satisfied in a non-contextual model. Thus, the winning probability in a non-contextual model is expected to be $(P_{w})_{ks}=5/6$.

By using the moment expansion of the joint probabilities \begin{eqnarray}
\nonumber
\label{moment}
 &&P(a,b,c|x,y,z)=\frac{1}{8}\Big[1+(-1)^a\langle A_{x}\rangle +(-1)^b\langle B_{y}\rangle \\
\nonumber
&+&(-1)^c\langle C_{z}\rangle+(-1)^{a\oplus_{2}b}\langle A_{x} B_{y}\rangle +(-1)^{b\oplus_{2}c}\langle B_{y}C_{z}\rangle\\
&+& (-1)^{a\oplus_{2}c}\langle A_{x} C_{z}\rangle+(-1)^{a\oplus_{2}b\oplus_{2}c}\langle A_{x} B_{y} C_{z}\rangle \Big]
\end{eqnarray} 
the winning probability in Eq. (\ref{sucprob1}) can be cast into the following form

\begin{align}
	\mathcal{P}_{w}=\frac{1}{2}\Big[1+\frac{\left\langle\Delta\right\rangle_{ks}}{6}\Big]
\end{align}
where $\langle\Delta\rangle$ is the Mermin's expression given by Eq. (\ref{newmc}). Clearly, in a non-contextual model $\left\langle\Delta\right\rangle_{ks}=4$ providing $(\mathcal{P}_{w})_{ks}=5/6$. In QM, for any state in two-qubit system and for the choices of observables in Eq. (\ref{pmarray}), we get $\Delta_{QM}=6$ providing $(\mathcal{P}_{w})_{QM}=1$.

This communication game may have independent importance elsewhere. Here we use it to demonstrate the certification of two-bit of randomness by KS quantum contextuality through the violation of Mermin's magic-square inequality. As already indicated earlier, for a qubit system by using the sequential measurements of non-commuting observables one can generate an arbitrary bit of randomness but that cannot be certified by quantum mechanics. In other words, such randomness can be reproduced by a classical model. Here we use a suitable set of six sequential measurements involving three commuting observables in a two-qubit system which enables the two-bit of certification of randomness in a semi-device independent way. 

In order to demonstrate this, our task should be to minimize the highest probability ( say, $G$) of the joint probabilities in sequential measurements which enable the violation of the non-contextuality inequality in Eq. (\ref{newmc}). Such certification of randomness thus requires solving the following optimization problem. 

\begin{eqnarray}
\label{max}
	&{\text {Minimize :}}   \ G={max}_{a,b,c} P(a,b,c|x,y,z)\\
	\nonumber
	&{\text {subject to:}}  \ \Delta_{QM}=\sum\limits_{x,y,z}\alpha_{x,y,z}^{a,b,c} P(a,b,c|x,y,z) \\
	\nonumber
	&{\text {with }} \ x=y=z; \ \ {\text{or}} \ y=x\oplus_{3}1,  z=x\oplus_{3} 2\end{eqnarray}
Here $	P(a,b,c|x,y,z)=Tr[B_{y}^{b} A_{x}^{a}\rho A_{x}^{a} B_{y}^{b} \ C_{z}^{c}]$ and $\alpha_{x,y,z}^{a,b,c}\in \{-1,+1\}$ is a function needs to be suitably chosen for different values of inputs and outputs. In the following, we argue that $G_{min}=1/4$ leading to the certification of two bit of randomness.    

It can be seen that $\Delta_{QM}=6$ requires each of the six three-measurement sequential correlations  

\begin{eqnarray}
	Tr[\rho A_x B_y C_z]&=&	P(a\oplus_{2}b\oplus_{2} c=0|x,y,z)\\
	\nonumber
	&-&P(a\oplus_{2}b\oplus_{2} c=1|x,y,z)=\pm 1
\end{eqnarray}
 In particular, $Tr[\rho A_x B_y C_z]$ equals to 1 if $x=y=z$, and -1 if $y=x\oplus_{3}1,  z=x\oplus_{3}$.  One of such choices of observables are provided in Eq.(\ref{pmarray}). Since all the two-qubit observables in Eq. (\ref{pmarray}) are dichotomic having eigenvalues $\pm 1$ then for satisfying  $Tr[\rho A_x B_y C_z]=\pm1$ the following conditions are required. i) For every $x,y,z$, each of the four probabilities $P(a\oplus_{2}b\oplus_{2} c=1|x,y,z)=0$ when $x=y=z$ and ii) each of the four probabilities $P(a\oplus_{2}b\oplus_{2} c=0|x,y,z)=0$ when $y=x\oplus_{3}1,  z=x\oplus_{3}$. 

Let us analyze the above two conditions. When $x=y=z$, from Eq. (\ref{pmarray})  we have $A_{x}B_{y}=C_{z}$, $B_{y}C_{z}=A_{x}$ and $A_{x} C_{z}=B_{y}$ along with $A_{x}B_{y}C_{z}=1$. In this case, the joint probabilities in Eq. (\ref{moment}) can be written as 
\begin{eqnarray}
\label{jp}
 &&P(a,b,c|x=y=z)=\frac{1}{8}\Big[1+(-1)^{a\oplus_{2}b\oplus_{2}c} \\
\nonumber
&+& \left((-1)^a +(-1)^{b\oplus_{2}c}\right) \langle A_{x}\rangle +\left((-1)^b +(-1)^{a\oplus_{2}c}\right) \langle B_{y}\rangle \\
\nonumber
&+&\left((-1)^c +(-1)^{a\oplus_{2}b}\right)\langle C_{z}\rangle \Big]
\end{eqnarray} 
Since $a\oplus_{2}b\oplus_{2} c=0$ implies $a=b\oplus_{2} c$ and so on,  we can write
\begin{eqnarray}
\label{jpp}
 &&P(a\oplus_{2}b\oplus_{2} c=0|x=y=z)\\
\nonumber
&=&\frac{1}{8}\Big[2+(-1)^a 2\langle A_{x}\rangle +(-1)^b 2\langle B_{y}\rangle +(-1)^c 2\langle C_{z}\rangle\Big]
\end{eqnarray} 
By noting that $a\oplus_{2}b\oplus_{2} c=1$ implies $a\oplus_{2}1=b\oplus_{2} c$, it is simple to show from Eq. (\ref{jp}) that $P(a\oplus_{2}b\oplus_{2} c=1|x=y=z)=0$. 

Similarly, when $y=x\oplus_{3}1,  z=x\oplus_{3} 2$ we have $A_{x}B_{y}=-C_{z}$, $B_{y}C_{z}=-A_{x}$ and $A_{x} C_{z}=-B_{y}$ along with $A_{x}B_{y}C_{z}=-1$. In this case the joint probabilities given by Eq. (\ref{moment}) can be written as
'\begin{eqnarray}
\label{jp1}
 &&P(a,b,c|x, y=x\oplus_{3}1,  z=x\oplus_{3} 2)=\\
\nonumber
&=&\frac{1}{8}\Big[1-(-1)^{a\oplus_{2}b\oplus_{2}c}  +\left((-1)^a  -(-1)^{b\oplus c}\right) \langle A_{x}\rangle   \\
\nonumber
&+&  \left((-1)^b  -(-1)^{a\oplus c}\right) \langle B_{y}\rangle+ \left((-1)^c  -(-1)^{a\oplus b}\right) \langle C_{z}\rangle \Big]
\end{eqnarray} 

It can be seen that when $a\oplus_{2}b\oplus_{2} c=1$, the joint probabilities 
\begin{align}
	P(a\oplus_{2}b\oplus_{2} c=1|x, y=x\oplus_{3}1,  z=x\oplus_{3} 2) \\
	\nonumber
	=P(a\oplus_{2}b\oplus_{2} c=0|x=y=z)
\end{align}
 as given by Eq. (\ref{jpp}) and when $a\oplus_{2}b\oplus_{2} c=0$ one has 
\begin{align}
	P(a\oplus_{2}b\oplus_{2} c=0|x, y=x\oplus_{3}1,  z=x\oplus_{3} 2) \\
	\nonumber
	=P(a\oplus_{2}b\oplus_{2} c=1|x=y=z)=0
\end{align}

Now, for example, if one considers $a=b=c=0$ and $\rho$ is a simultaneous eigenstate of the three commuting observables $A_{1}$, $B_{1}$ and $C_{1}$ having eigenvalue $1$, then from Eq. (\ref{jpp}), we find $P(000|0,0,0)=1$. This in turn provides $G=1$,  leading to zero randomness. However, it is the extreme case. But, for a maximally mixed state in two-qubit system $\rho=\mathbb{I}_{4}/4$,  for $a,b,c \in \{0,1\}$ and $x,y,z\in \{1,2,3\}$,  from Eq. (\ref{jpp}) we have

\begin{align}
	P(a\oplus_{2}b\oplus_{2} c=1|x, y=x\oplus_{3}1,  z=x\oplus_{3} 2) \\
	\nonumber
	=P(a\oplus_{2}b\oplus_{2} c=0|x=y=z)=\frac{1}{4}
\end{align}
 In such a case, from Eq. (\ref{max}) we have $G=1/4$  and as already mentioned $\langle\Delta\rangle_{Q}=6$. Thus, two-bit of randomness can be certified in our semi-device independent prepare-measure scenario. The novelty of our scheme is that this semi-device independent prepare-measure game can be directly linked to Mermin's magic square proof of contextuality. Such a randomness generation protocol has not been hitherto introduced in the literature.  

In the following, we provide an argument that the magic-square proof can also be cast as a proof of preparation contextuality which illustrates the fact that although a maximally mixed state prepared by many distinct procedures (various convex combinations of pure states) cannot be distinguished in quantum theory. But the probability distribution of the ontic states in an ontological model corresponding to the same mixed state prepared by different procedures can be distinct. This is, in fact, the notion of preparation contextuality.        

\section{Magic-square game as a preparation contextual game}

 For our purpose, we first provide an analogy by noting a prepare-measure communication game widely known as parity-oblivious random access code \cite{spek09}. The simplest version of it, the $2\rightarrow 1$ case is the following. Alice holds a $2$-bit string $i$  chosen uniformly at random from $\{00,01,10,11\}$. Bob can choose any bit $j=\{1,2\}$ to recover $i_{j}$ bit of Alice with a probability. The condition of the task is that Bob's output must be the bit $b =i_{j}$ i.e., the $j^{th}$ bit of Alice's input string $i$ with the constraint that no information about any parity $(i_1 \oplus_{2} i_2)$ of input string $i$ can be transmitted to Bob. This means Alice posses two sets; even-parity set includes $\{00,11\}$ and odd-parity set includes $\{01,10\}$. In quantum theory, Alice encodes her $2$-bit string of $i$ into pure states $\rho_{i}$ prepared by a procedure $P_i$ so that the parity oblivious condition is satisfied. This quantum mechanically means $(1/2)\rho_{00}+(1/2)\rho_{11}=\mathbb{I}/2$ and $(1/2)\rho_{10}+(1/2)\rho_{01}=\mathbb{I}/2$.  The success probability of parity-oblivious random access code in classical theory is $0.75$ but in quantum theory $0.85$ \cite{spek09,pan18}. 

In the existing randomness certification using prepare-measure random access code \cite{li11,li12}, the notion of preparation contextuality was not mentioned. Let us note a crucial point relevant to our work. Alice can steer four encoding quantum states for Bob by measuring two observables on a pre-shared two-qubit entangled state between Alice and Bob. Alternatively, she can prepare those states by taking a maximally mixed state $\rho=\mathbb{I}/2$ and randomly performing the measurements of two observables $M_1$ (having eigenstates $\rho_{00}$ and $\rho_{11}$) and $M_2$ (having eigenstates $\rho_{01}$ and $\rho_{10}$). Since these two preparation procedures (say, P and $P^{\prime}$) satisfy the parity oblivious condition in quantum theory,  then equivalently in an ontological model, we have preparation non-contextuality assumption, i.e.,  $\mu_{P}(\lambda|\mathbb{I}/2)=\mu_{P^{\prime}}(\lambda|\mathbb{I}/2)$. It is argued \cite{spek09} that the quantum supremacy in prepare-measure random access code appears due to the preparation contextuality. This argument was generalized for $n\rightarrow 1$ parity-oblivious random access code in \cite{asmita}. By keeping this scenario in mind we provide a sketch of how one can cast the magic-square game as a preparation contextual game.

For this, consider once again tour prepare-measure game where Alice, by using suitable sequential measurement on a given state $\rho$, prepares the states $\rho_{ab}$ for Charlie. He randomly performs three measurements $C_1$, $C_2$ and $C_3$. The rule of the game remains the same. In particular, when Alice sequentially chose $A_{1}$ and $B_{1}$ (or $A_{2}$ and $B_{3}$) respectively, then Charlie chooses $C_{1}$ and similarly for other such cases. Let the initial state is $\rho=\mathbb{I}_{4}/4 \in \mathbb{C}^{2}\otimes \mathbb{C}^{2}$. By using a preparation procedure $P_{1}$, Alice sequentially perform measurements of commuting observables $A_{1}$ and $B_{1}$ on $\rho$ for preparing four pure states (which are the common eigenstates of $A_{1}$ and $B_{1}$) for Charlie, with equal probability. Hence, the maximally mixed state remains the same but having different compositions of pure states.  Similarly,  consider a preparation procedure $P_{2}$ where sequential measurements of $A_{2}$ and $B_{3}$ are performed by Alice. This again leads to maximally mixed state $\mathbb{I}_{4}/4$ but with another composition corresponding to the common eigenstates of $A_2$ and $B_3$.  In quantum theory,  $\mathbb{I}_{4}/4$ prepared by two preparation $P_{1}$ and $P_{2}$ cannot be distinguished by any measurement and equivalently leads to the assumption of preparation non-contextuality in an ontological model. Symbolically, $\mu_{P_{1}}(\lambda|\mathbb{I}_{4}/4)=\mu_{P_{2}}(\lambda|\mathbb{I}_{4}/4)$. Along the same line of argument for the remaining two values of $x$ and $y$ each, we can say there is four more such preparation procedure for preparing $\mathbb{I}_{4}/4$ and preparation non-contextuality can be assumed for them as well. Since the rule of the game remains the same, the success probability in a preparation non-contextual model is $5/6$ and in quantum theory $1$. Thus, only a preparation contextual ontological model can reproduce quantum success probability.  

This is not very surprising. There exists proof that demonstrates that the preparation non-contextuality implies KS non-contextuality \cite{leifer}. The importance of the above argument lies in the fact that although the same maximally mixed state prepared by distinct preparations \emph{cannot} be distinguished by any measurement in quantum theory but one cannot extrapolate that kind of equivalence in an ontological model. In other words, probability distributions of ontic states corresponding to the same density matrix prepared by different preparation procedures cannot be represented as preparation non-contextually in an ontological model. The above discussion thus illustrates another possible form of non-classicality in the magic-square game is the preparation contextuality.

\section{Magic-star proof and randomness certification}
Next, we consider the magic-star proof of Mermin which requires minimum three-qubit system to run the argument.
\begin{figure}[ht]
\centering
\label{fig}
\includegraphics[width=1\linewidth]{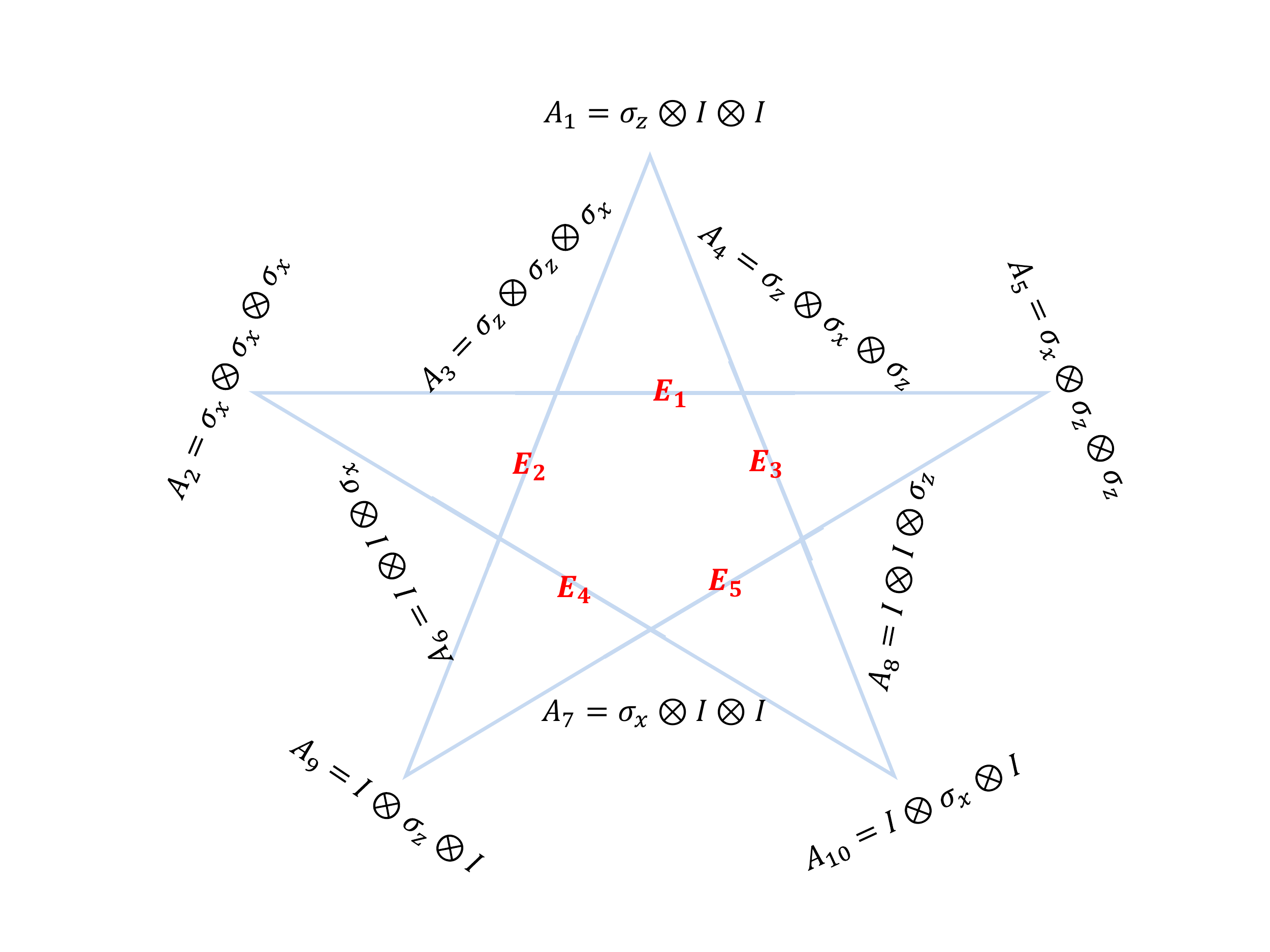}
\caption{Magic-star proof diagram}
\end{figure}
In this proof, ten three-qubit observables are used as in Fig. 1 and they are so arranged that each of the five edges (denoted as $E_j$ with $j=1,2,3,4,5$) contains four mutually commuting observables. For any state in three-qubit system, $Tr[\rho E_1]=Tr[\rho(A_2 A_3 A_4 A_5)=-1$ and rest of the other edges $Tr[E_{j}\rho]=1$ where $j\neq 1$. In a KS non-contextual model one may then write, $\xi(E_1)=\xi(A_2)\xi( A_3)\xi( A_4)\xi( A_5)=-1$ and similarly for others. Following the magic-square proof, one can easily find a contradiction of KS non-contextuality. However, for experimental testing the KS-type logical argument, inequality based proofs are required. An inequality based on magic-star proof can be written as
\begin{eqnarray}
	\mathcal{\beta}_{ks}=-\langle E_1\rangle + \langle E_2\rangle+\langle E_3 \rangle+ \langle E_4\rangle+ \langle E_5\rangle \leq 3
\end{eqnarray}
In quantum mechanics, $\mathcal{\beta}_{Q}=5$ violating the KS non-contextual bound. This can only be achieved, if $\langle E_1\rangle=-1$ and other  four sequential correlations provide $+1$. From the construction of magic star in Fig.1, one can find that for three-qubit system such choice of observables are found. 
Following the approach used above, one may minimize the maximum joint probability so that
\begin{eqnarray}
	{\text {Minimize :}} && \ G={max}_{e_j} P(e_j|E_j)\\
	\nonumber
	&&{\text {subject to:}} \ \ \mathcal{\beta}=\sum\limits_{j} \alpha_{j}^{e_j} P(e_j|E_{j}) \\
\end{eqnarray}
where $e_1= (a_1,a_2, a_3,a_4)\in \{0,1\}$ and similarly for others. Here $\alpha_{j}^{e_j}$ takes $\pm 1$ value and 
\begin{align}
	P(e_1|E_1)=Tr[A_{3}^{a_3}A_{2}^{a_2}A_{1}^{a_1}\rho A_{1}^{a_1}  A_{2}^{a_2} A_{3}^{a_3} A_{4}^{a_4} ]
\end{align}
As mentioned above, for maximum value $\langle\beta\rangle_{QM}=5$, we have $\langle E_{1}\rangle=-1$ and rest $\langle E_{j\neq 1}\rangle$ are $+1$. Now, to obtain $\langle E_{1}\rangle_{\rho}= P(a_2\oplus_{2}a_3\oplus_{2}a_4 \oplus_{2} a_5=0|E_2)-P(a_2\oplus_{2}a_3\oplus_{2}a_4 \oplus_{2} a_5=1|E_2)=-1$, we have $P(a_2\oplus_{2}a_3\oplus_{2}a_4 \oplus_{2} a_5=0|E_2)=0$. Among other eight probabilities $P(a_2\oplus_{2}a_3\oplus_{2}a_4 \oplus_{2} a_5=1|E_2)$, there are many possibilities to add up to $1$. But for a maximally mixed state in three-qubit system, we have all the eight joint probabilities are $1/8$. This in tun provides the certification of three bit of randomness using the Mermin's magic-star proof of KS contextuality.

\section{Summary and Discussion}
To summarize, we provided two interesting semi-device independent randomness generation schemes certified by KS contextuality. For our purpose, we used two KS proofs of Mermin \cite{mermin}, viz., the magic-square, and the magic-star proofs for demonstrating the generation of two and three-bit of randomness. Specifically, we have cast the Mermin's proof KS theorem (involving sequential measurements of three or more commuting observables in a bounded dimension) as a prepare-measure communication game which sufficed our purpose here.  Note that, for certifying two or more bit of randomness in a two-qubit entangled system, one requires unsharp measurements \cite{acin16,and18,curchod}. On the other hand, in our protocols, the sharp sequential measurements of commuting observables in single two-qubit and three-qubit systems are used for certifying two- and three-bit of randomness respectively. 

Recently, a couple of experimental studies have been reported to demonstrate the randomness generation protocols using quantum contextuality. In \cite{um}, random bits are certified based on contextuality tests through the violation of the Klyachko-Can-Binicioglu-Shumovsky  inequality \cite{kly} using single qutrit. In another experiment, Kulikov \emph{et al.,}  \cite{kulikov} generated quantum number for qutrit system certified by a stronger form KS contextuality based on the proposal in \cite{abbott}. Our protocol requires maximally mixed state in two(or three)-qubit system and the certify the higher random bits compared to \cite{um,kulikov}.

The device-independent randomness certification through the violation of a Bell's inequality requires a suitable entangled state. Strictly speaking, a protocol based on Bell's inequality is in fact randomness expansion protocol as prior input randomness is required for Bell test. Also, the loophole-free test of Bell's inequality has to be guaranteed. Although some recent experiments \cite{loopholefree} has achieved this goal but it suffers from a low bit generation rate and the ratio of output-input randomness more than one still remains a challenge in the practical experiments. The semi-device independent protocols in prepare-measure scenario \cite{li11,li12,Wen,pas,van,brask,ioa} are free from other drawbacks arises with the Bell test. We note again that our protocols are also semi-device independent which do not use the entanglement and certification is provided by KS quantum contextuality. We have used only a couple of general assumptions common to any randomness certification protocol in prepare-measure scenario; the dimension is bounded and there is no information leakage form the devices of Alice and Charlie.  

We note here that there are different kinds of two-qubit systems that have already been realized in various experiments. For example, the path and the spin degrees of freedom of single neutrons that was used Hasegawa \emph{et. al.} \cite{hasegawa}, path and polarization degrees of freedom of a single photon \cite{liu,ams} and a pair of trapped ions that is experimentally realized by Kirchmair \emph{et. al}\cite{nature}. Hence, the schemes proposed here can be experimentally tested using existing technologies. Finally, we conjecture that for suitable Mermin-type proof of KS contextuality in a higher dimensional system, any arbitrary bit of randomness can be generated. This calls for further study.  

\acknowledgments
 AKP acknowledges the support from the project DST/ICPS/QuEST/2018/Q-42.

\end{document}